\documentclass[aps,prl,twocolumn,amsmath,amssymb,groupedaddress]{revtex4}
\usepackage{epsfig}
\begin{document}

\title{Light scattering and phase behavior of Lysozyme-PEG
mixtures}

\author{J. Bloustine$^{1}$}
\author{T. Virmani$^{1}$} \thanks{Present Address: Center
for Basic Neuroscience, The University of Texas Southwestern
Medical Center, Dallas, TX 75390}
\author{G.~M. Thurston$^{2}$}
\author{S. Fraden$^{1}$} \affiliation{$^1$Complex Fluids Group,
Martin Fisher School of
Physics, Brandeis University, Waltham, MA 02454 \\
$^2$Department of Physics, Rochester Institute of Technology,
Rochester, NY 14623}
\date{\today}


\begin{abstract} Measurements of liquid-liquid phase
transition temperatures (cloud points) of mixtures of a protein
(lysozyme) and a polymer, poly(ethylene glycol) (PEG) show that
the addition of low molecular weight PEG stabilizes the mixture
whereas high molecular weight PEG was destabilizing. We
demonstrate that this behavior is inconsistent with an entropic
depletion interaction between lysozyme and PEG and suggest that an
energetic attraction between lysozyme and PEG is responsible. In
order to independently characterize the lysozyme/PEG interactions,
light scattering experiments on the same mixtures were performed
to measure second and third virial coefficients. These
measurements indicate that PEG induces repulsion between lysozyme
molecules, contrary to the depletion prediction. Furthermore, it
is shown that third virial terms must be included in the mixture's
free energy in order to qualitatively capture our cloud point and
light scattering data. The light scattering results were
consistent with the cloud point measurements and indicate that
attractions exist between lysozyme and PEG.
\end{abstract}
\maketitle


The addition of small polymers to a dispersion of large colloids
can result in precipitation or crystallization of the colloids.
Even when the sole colloid/polymer interaction is steric
repulsion, an attraction between a pair of colloids is generated
by the exclusion of polymer molecules from the region between the
colloids. This entropic effect is known as depletion attraction
\cite{asakura58,gast83}. The water soluble polymer poly(ethylene
glycol)(PEG) has been utilized extensively to induce protein
crystallization \cite{mcpherson99}. Can the mechanism of PEG
induced crystallization be explained by the purely entropic
depletion effect? This work addresses the related question of
whether or not PEG induces attraction between lysozyme molecules.

In addition to a liquid to crystal transition, protein solutions
exhibit a metastable liquid-liquid phase transition when cooled
\cite{Ishimoto77,taratuta90,muschol97,thomson87,annunziata03}.
This phase transition temperature is termed the \textit{cloud
point} ($T_{\mbox{\scriptsize{cloud}}}$) since at this temperature
a transparent protein solution becomes turbid as liquid droplets
of high protein concentration form in a liquid of lower protein
concentration. The effect of added PEG on
$T_{\mbox{\scriptsize{cloud}}}$ was studied for several of the
$\gamma$-crystallin proteins by Benedek and co-workers who found
that $T_{\mbox{\scriptsize{cloud}}}$ of $\gamma$S crystallin
increases as PEG is added \cite{annunziata03} in agreement with
depletion attraction whereas for $\gamma$D crystallin their
results \cite{benedek02} indicate a departure from pure depletion.
Galkin and Vekilov \cite{galkin00} studied lysozyme/PEG mixtures
and found that the effect of PEG molecular weight on
$T_{\mbox{\scriptsize{cloud}}}$ depended on the solution's ionic
strength so no clear evaluation of the depletion effect could be
made. Because the cloud point depends on salt type and
concentration~\cite{broide96,grigsby01}, all our measurements were
performed in the same solution conditions.

We model the thermodynamics of a protein/polymer mixture by
expanding the excess Gibbs free energy ($G$) of a two component
solution relative to that of the solvent in powers of the
densities of the two independent solutes, labelled $1$ and $2$.
\begin{eqnarray}\label{eq:F-virial-expand}
g=\frac{G}{V k_B T}=\rho_1\ln \rho_1+\rho_2\ln \rho_2+
B_{11}\rho_1^2+2B_{12}\rho_1\rho_2\nonumber
\\
+B_{22}\rho_2^2+ C_{111}\rho_1^3+3C_{112}\rho_1^2\rho_2+\ldots
\end{eqnarray}
In Eq.(\ref{eq:F-virial-expand}) $\rho_i=N_i/V$ [Volume$^{-1}$];
$k_B$ is Boltzmann's constant; $T$ is the absolute temperature;
$B_{ij}$ [Volume] are the second virial coefficients and $C_{ijk}$
[Volume$^2$] are the third virial coefficients. Virial
coefficients are related to integrals of the potential of mean
force between molecules and depend on $T$ \cite{hil60}.

A bidisperse hard sphere mixture was used as the reference system
for lysozyme/PEG mixtures because hard sphere systems provide a
natural scale for virial coefficients. Hard spheres cannot
interpenetrate, but have no other interactions. The virial
coefficients in Eq.(\ref{eq:F-virial-expand}) for a bidisperse
hard sphere mixture are \cite{lebowitz64}:
 \begin{equation}\label{eq:secondvirials-hs}
 B_{11}^{\mbox{\scriptsize{HS}}}=(16\pi/3)r_1^3, \; \; \; \;  \; \; B_{12}^{\mbox{\scriptsize{HS}}} = \frac{2\pi}{3}(r_1+r_2)^3,\\
\end{equation}
\begin{equation}\label{eq:thirdvirials-hs}
 C_{111}^{\mbox{\scriptsize{HS}}}=\frac{5}{16}(B_{11}^{\mbox{\scriptsize{HS}}})^2, \; \;  \; \; C_{112}^{\mbox{\scriptsize{HS}}} = \frac{8 \pi^2}{27}r_1^3(r_1^3+6r_1^2r_2+15r_1 r_2^2 +8r_2^3)
 \end{equation}
where $r_i$ are the hard sphere radii.

The ratio of lysozyme to PEG measured hydrodynamic radii, $r_H$,
in our experiments varied in the range $3 \geq
r_{\mbox{\scriptsize{H}}}^{\mbox{\scriptsize{lys}}}/r_{\mbox{\scriptsize{H}}}^{\mbox{\scriptsize{PEG}}}
\geq 0.8$. Therefore we need to account for the fact that PEG
molecules, which are nearly as large or larger than lysozyme
molecules, are not spherical and wrap partially around proteins thereby reducing the
polymer/protein excluded volume. This is done by defining an ideal
effective polymer radius ($r_{\mbox{\scriptsize{eff}}}$) which is
smaller than the polymer's radius of gyration $r_g$. Eisenriegler
\textit{et al}. \cite{eisenriegler96} found a closed formula for
computing $r_{\mbox{\scriptsize{eff}}}$ from the protein radius
and $r_g$ (see Fig.6 in ref. \cite{eisenriegler96}). Our procedure
was to equate the protein's radius with its measured $r_H$ whereas
the process to determine the PEG's $r_g$ is described in the
following paragraph.

 The PEG radii of gyration for the molecular weights employed here were
too small to measure with static light scattering because $r_g$ is
much less than the wavelength of light. In order to determine
$r_g$, we therefore first measured $r_H$ and then used the
relation $r_g/r_H=1.48M_2^{0.012}$ \cite{devanand91} where $M_2$
[g mol$^{-1}$] is the PEG molecular weight. This should be
compared to the theoretical value, $r_g/r_H=1.56$, for a polymer
in a good solvent \cite{teraoka02}. For lysozyme we found:
 $r_H^{\mbox{\scriptsize{lys}}}=2.2$~nm which falls within
the range of previously reported values \cite{grigsby00}. We
measured $r_H^{\mbox{\scriptsize{PEG}}}= $0.75 and 2.7 nm for PEG1k
and 8k respectively where the manufacturer's stated value for
$M_2$ is designated with the nomenclature PEGnk, meaning PEG
n$\times 10^3$ [g mol$^{-1}$]. The procedure outlined above yields
$r_{\mbox{\scriptsize{eff}}}=$1.2 and 3.8 nm for PEG1k and 8k.

We measured the cloud point temperature of lysozyme solutions as
an approximation of the spinodal decomposition temperature since
the cloud point closely tracks the spinodal \cite{thomson87}.
Starting from the free energy, we calculate how the spinodal
temperature changes with added polymer concentration at fixed
protein concentration. In all that follows the subscript 1 refers
to lysozyme and the subscript 2 refers to PEG. A two component
mixture at fixed concentration undergoes spinodal decomposition
when the temperature reaches $\mbox{T}_{\mbox{\scriptsize{sp}}}$
defined by \cite{defay54}:
\begin{equation*}
f(\mbox{T}_{\mbox{\scriptsize{sp}}},\rho_1,\rho_2)=\frac{\partial^2g}{\partial
\rho_1^2}\frac{\partial^2g}{\partial
\rho_2^2}-(\frac{\partial^2g}{\partial \rho_1 \partial
\rho_2})^2=0
\end{equation*}
 Imposing the constraints that the solution remains on the
spinodal curve when polymer is added and that the protein
concentration is constant one finds $\partial
\mbox{T}_{\mbox{\scriptsize{sp}}}/\partial \rho_2=-(\partial
f/\partial \rho_2)/(\partial f/\partial T)$. From
Eq.(\ref{eq:F-virial-expand}), one then obtains to first order in
$\rho_1$:
\begin{equation}\label{eq:dTdp-virials}
\lim_{\rho_2 \rightarrow 0}\frac{\partial
\mbox{T}_{\mbox{\scriptsize{sp}}}}{\partial
\rho_2}=\frac{-3C_{112}+2(B_{12}+3\rho_1 C_{112})^2}{\partial
B_{11}/\partial T+3\rho_1 \frac{\partial C_{111}}{\partial T}}
\end{equation}
In all that follows we have set $\partial C_{111}/\partial T =0$
since experimentally $C_{111}=0$ at all temperatures.

 We define the numerator of
Eq.(\ref{eq:dTdp-virials}) to be $ \gamma$. The PEG molecular
weight dependence enters Eq.(\ref{eq:dTdp-virials}) only through
the virial coefficients. From Eqs.(\ref{eq:secondvirials-hs},
\ref{eq:thirdvirials-hs}, \ref{eq:dTdp-virials}), it can be shown
that for a hard sphere mixture $\gamma>0$ for all sphere sizes
$r_1$ and $r_2$ and concentrations $\rho_1$. In the limit of small
polymers ($q=r_2/r_1\ll1$) we find that
$2\gamma/(B_{11}^{\mbox{\scriptsize{HS}}}/4)^2=12q^3$ in agreement
with scaled particle depletion models \cite{benedek02}.

In order to explore the dependence of
$T_{\mbox{\scriptsize{cloud}}}$ on PEG molecular weight further,
virial coefficients were determined from light scattering
experiments on the same lysozyme/PEG mixtures and then compared
with predictions of the depletion theory. Kirkwood \& Goldberg
\cite{kirkwood50} showed that the excess light scattering of a two
solute system ($R_{1+2}$) over that of a single solute system
($R_2$) can be written as:
\begin{equation} \label{eq:LS-2comp-short}
\frac{Kc_{1}}{R_{1+2}-R_{2}}=\alpha+\beta \times c_{1}
\end{equation}
Here $K=2(\pi n_o n_1)^2/N_A \lambda^4$ where $n_o$ is the solvent
refractive index, $n_i=\frac{d n}{d c_i}$ is the refractive index
increment of solute $i$, $N_A$ is Avogadro's number, $\lambda$ is
the wavelength of the incident radiation in vacuum, $c$[g
mL$^{-1}$] is the solute weight concentration and $R$ is the
Rayleigh ratio.

$\alpha$ and $\beta$ depend on the added polymer properties
($M_2$, $n_2$), concentration $c_2$, and the protein/polymer
interaction \cite{kirkwood50}:
\begin{equation} \label{eq:2comp-Int}
\alpha=\frac{1}{M_1}+c_2\frac{4M_{2}n_{2}}{M_{1}n_{1}}B_{12}
\end{equation}
 $B_{12}$ is
obtained from measurements of $\alpha$ as a function of polymer
concentration $c_2$ shown in Fig.\ref{fig:Int-All}.

The coefficient $\beta$ is given by
\begin{eqnarray}
& \beta  =  2 B_{11}+\nonumber \\
 & 2c_2 \left[(3C_{112}-2B_{12}^2 M_2)+
\frac{2(3C_{112}+2B_{11}B_{12}M_1)M_2 n_2}{M_1 n_1} \right]
 \label{eq:2comp-Slope}
\end{eqnarray}
$C_{112}$ is obtained from measurements of $\beta$ as a function
of $c_2$, as shown in Fig. \ref{fig:Int-All}, since all the other
quantities in Eq.(\ref{eq:2comp-Slope}) are determined
independently.

 It is possible to view the
two component polymer/protein solution as an effective one
component protein solution. The effective protein/protein second
virial coefficient, $B_{11}^{\mbox{\scriptsize{eff}}}$, can be
obtained from Eq.(\ref{eq:2comp-Slope}) by imagining the addition
of invisible polymers ($n_2=0$) to the protein solution
\cite{vrij76,tong90} which yields:
$B_{11}^{\mbox{\scriptsize{eff}}}=B_{11}+c_2 \left
[(3C_{112}-2B_{12}^2 M_2)\right]$. Experimentally we do not index
match the polymer $(n_2 \neq 0)$. Instead we measure the virial
coefficients and calculate $B_{11}^{\mbox{\scriptsize{eff}}}$.

Hen egg white lysozyme was purchased from Seikagaku America. $M_1$
based on sequence is 14,400 g mol$^{-1}$. Poly(ethylene glycol)
(PEG) was purchased from Sigma and Fluka.

The protein and PEG were dissolved in a 0.2M sodium phosphate
buffer with NaCl 0.5M at pH 6.2 where M $\equiv$ [mol L$^{-1}$].
All solutions were centrifuged at $\sim 12000\times g$ for 1 hour
and then passed through 0.2$\mu$m filters directly into precleaned
scattering cuvettes. The lysozyme concentrations were measured by
UV absorption using an extinction coefficient
$\epsilon_{280\mbox{\scriptsize{nm}}}=2.64$ mL mg$^{-1}$
cm$^{-1}$.

 Refractive index increments
of lysozyme and PEG were measured using a Brookhaven Instruments
 differential refractometer at
$\lambda= $ 620~nm. For lysozyme $d n/d c= 1.85 \times 10^{-4}$ mL
mg$^{-1}$ and for PEG $d n/d c= 1.34 \times 10^{-4}$ mL mg$^{-1}$
independent of $M_2$ and $T$.

Cloud point temperatures were determined by optical microscopy.
Rectangular glass capillaries (0.1 mm path length, VitroCom) were
filled with solution, flame sealed and then placed in a custom
built temperature controlled microscope stage. The temperature at
which homogeneous nucleation of dense droplets occurred was called
$T_{\mbox{\scriptsize{cloud}}}$. The temperature of each solution
was cycled up and down at approximately 1$^\circ$C min$^{-1}$
through $T_{\mbox{\scriptsize{cloud}}}$ several times for each
measurement. Little or no difference ($\leq0.5^{\circ}$C) was
observed when comparing $T_{\mbox{\scriptsize{cloud}}}$ obtained
by cooling and heating.

 The static and dynamic light scattering experiments (SLS \& DLS)
were performed using an ALV goniometer and correlator system
in the \textit{vu} polarization mode. Absolute Rayleigh ratios of
aqueous solutions were determined by using pure toluene as a
standard whose Rayleigh ratio is known
\cite{kaye73}. The lysozyme and PEG hydrodynamic radii
$r_{\mbox{\scriptsize{H}}}$ were obtained from DLS measurements on
dilute solutions \cite{bern76}.

\begin{figure}
\epsfig{file=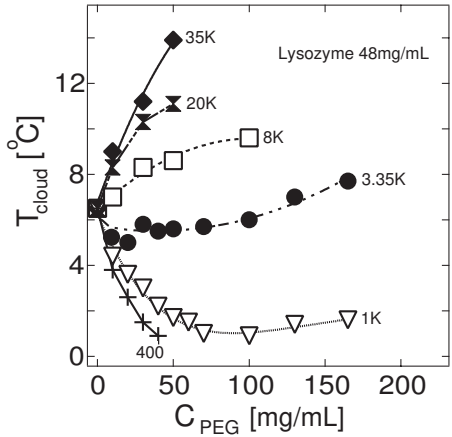,width=6cm}
 \caption{Cloud
point temperatures ($\mbox{T}_{\mbox{\scriptsize{cloud}}}$) of
lysozyme/PEG mixtures as a function of PEG concentration for
different PEG molecular weights ranging from 400 to 35k g/mol at a
fixed lysozyme concentration of 48 mg/mL in the phosphate buffer.
The lines are guides to the eye.} \label{fig:lys-Tc-vs-PEG}
\end{figure}

\begin{figure}
\epsfig{file=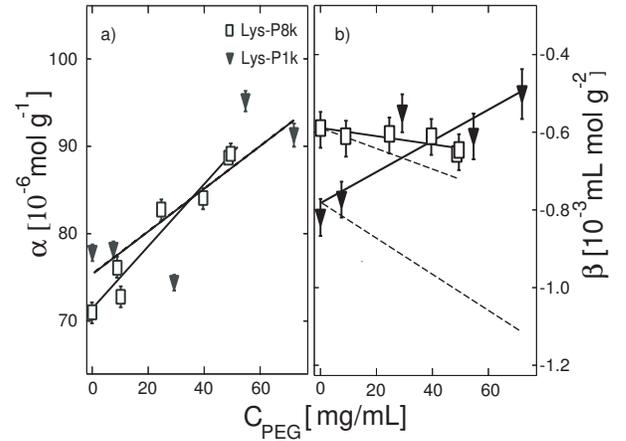,width=8cm}
 \caption{The
variation of $\alpha$ (Eq.(\ref{eq:2comp-Int})) and $\beta$
(Eq.(\ref{eq:2comp-Slope})) on PEG concentration ($c_2$) at
T=30$^{\circ}$C is shown for lysozyme/PEG1k and PEG8k mixtures
dissolved in the phosphate buffer. The solid lines in panel a)
indicate the fits to $\alpha $ used to obtain $B_{12}$ by
Eq.(\ref{eq:2comp-Int}). The solid lines in panel b) indicate the
fits to $\beta$ used to obtain $C_{112}$ by
Eq.(\ref{eq:2comp-Slope}). The dashed lines in panel b) show the
dependence of $\beta$ on PEG concentration assuming $C_{112}=0$.}
\label{fig:Int-All}
\end{figure}

Our $T_{\mbox{\scriptsize{cloud}}}$ measurements of lysozyme/PEG
mixtures at constant ionic strength reveal a systematic trend:
whereas $T_{\mbox{\scriptsize{cloud}}}$ increases upon the
addition of high molecular weight PEG, it \textit{decreases} for
low molecular weight PEG as shown in Figure
\ref{fig:lys-Tc-vs-PEG}. We extracted $\lim_{\rho_2 \rightarrow
0}\partial T_{\mbox{\scriptsize{cloud}}}/\partial \rho_2$ from
this data. Measurements of $T_{\mbox{\scriptsize{cloud}}}$ made at
higher lysozyme concentrations near lysozyme's critical point
showed that $\lim_{\rho_2 \rightarrow 0}\partial T/\partial
\rho_2$ is independent of lysozyme concentration (data not shown).
The hard sphere mixture model predicts that the cloud point
temperature dependence on PEG concentration cannot change sign
with PEG molecular weight, i.e. $\gamma>0$. However, our
experiments reveal qualitatively different behavior. For dilute
PEG concentrations, low molecular weight PEGs depress the cloud
point, i.e. \emph{stabilize} the solution, which is opposite to
the depletion prediction, while higher molecular weight PEGs raise
the cloud point, i.e. destabilize the solution.

Table \ref{tab:cross-virials} shows the PEG molecular weights
obtained from SLS data. PEG1k and 8k second virial coefficients as
functions of temperature were found to be
$B_{22}^{\mbox{\scriptsize{P1k}}}(T)[\mbox{mL mol
\,g}^{-2}]=0.01-(8.8\times10^{-5} \times T[^{\circ}\mbox{C}])$ and
$B_{22}^{\mbox{\scriptsize{P8k}}}(T)[\mbox{mL mol
\,g}^{-2}]=5.6\times10^{-3}-(9.5\times 10^{-5} \times
T[^{\circ}\mbox{C}])$, which agree well with previous results
\cite{devanand91}. Table \ref{tab:cross-virials} shows that
measured PEG second virial coefficients are the same order as
those for equivalent hard spheres, implying
 that PEG interactions are repulsive. For
lysozyme: $B_{11}(T=20^{\circ}\mbox{C})=-3.95\pm0.3\times10^{-4}$
mL mol g$^{-1}$, $\partial B_{11}/\partial T=1\times10^{-5}$ mL
mol g$^{-2}$ $^{\circ}$C$^{-1}$ and $M_{1}=13800\pm500$ g
mol$^{-1}$.

Eq.(\ref{eq:2comp-Int}) demonstrates that important information
about lysozyme/PEG interactions is contained in $\alpha(c_2)$,
namely $B_{12}$. The measured values of $B_{12}$ shown in Table
\ref{tab:cross-virials} are consistent with previous measurements
\cite{king88,atha81}. If interference between the scattering from
lysozyme and PEG was negligible then $\partial \alpha/\partial
c_2=0$, as seen from Eq.(\ref{eq:2comp-Int}) with $(M_2 n_2)/(M_1
n_1)=0$. However, Figure \ref{fig:Int-All} panel a) shows that
$\partial \alpha/\partial c_2>0$ consistent with Prausnitz and
co-workers \cite{king88}. This demonstrates that interference from
lysozyme and PEG scattering may not be ignored as done by Kulkarni
\textit{et al}. \cite{kulk99} and that $\alpha$ may not be treated
as a constant. Figure \ref{fig:Int-All} panel b) displays the
variation of $\beta$ with PEG concentration. The solid lines
through the data points are the linear fits used to obtain the
values of $C_{112}$ by Eq.(\ref{eq:2comp-Slope}) shown in Table
\ref{tab:cross-virials}. We found $B_{12}$ and $C_{112}$ to be
temperature independent.

Much interest has been generated by the conjecture that the second
virial coefficient may be sufficient to predict protein solution
phase behavior \cite{haas98,vliegenthart00}. The measured
dependencies of $T_{\mbox{\scriptsize{cloud}}}$ and $\beta$ on PEG
concentration show that third virial coefficients must be included
in the free energy of lysozyme/PEG mixtures, i.e. $C_{112}\neq0$.
If $C_{112}=0$ then Eq.(\ref{eq:2comp-Slope}) predicts
$\beta(c_2)$ shown as the dashed lines in panel b) of
Fig.\ref{fig:Int-All}, which disagree with our data. Additionally,
if $C_{112}=0$ then Eq.(\ref{eq:dTdp-virials}) predicts that
$\partial \mbox{T}_{\mbox{\scriptsize{sp}}}/\partial \rho_2$
cannot change sign as a function of PEG molecular weight, which is
inconsistent with Fig.\ref{fig:lys-Tc-vs-PEG}.

 Table
\ref{tab:cross-virials} compares our measured values of the mixed
virial coefficients, $B_{12}$ and $C_{112}$, to
$B_{12}^{\mbox{\scriptsize{HS}}}$ and
$C_{112}^{\mbox{\scriptsize{HS}}}$, calculated from
Eqs.(\ref{eq:secondvirials-hs}, \ref{eq:thirdvirials-hs}). The
lysozyme equivalent hard sphere radius was taken to be
$r_{\mbox{\scriptsize{H}}}$ and the PEG equivalent hard sphere
radii were taken to be $r_{\mbox{\scriptsize{eff}}}$. The
lysozyme/PEG mixed virial coefficients
 are less than those of equivalent hard sphere mixtures. Therefore, attractive
 interactions must exist between the protein
 and polymer or the repulsion must be less than that between
 hard spheres.

The measured virial coefficients allow us to calculate the
variation of the effective lysozyme second virial coefficient with
PEG concentration, $\partial
B_{11}^{\mbox{\scriptsize{eff}}}/\partial c_2$. Depletion theory
predicts that PEG induces attraction between lysozyme molecules in
which case $\partial B_{11}^{\mbox{\scriptsize{eff}}}/\partial
c_2<0$ as seen from
Eqs.(\ref{eq:secondvirials-hs},\ref{eq:thirdvirials-hs}).
Conversely, if $\partial B_{11}^{\mbox{\scriptsize{eff}}}/\partial
c_2>0$ then PEG induces repulsion between lysozyme molecules. For
PEG 8k we find $\partial B_{11}^{\mbox{\scriptsize{eff}}}/\partial
c_2=0.33\pm0.2\times10^{-3}$ mL$^{2}$ mol g$^{-3}$
 and for PEG 1k we
find that $\partial B_{11}^{\mbox{\scriptsize{eff}}}/\partial
c_2=1.6\pm0.2\times10^{-3}$ mL$^{2}$ mol g$^{-3}$. We conclude
that adding PEG weakens the attraction between lysozyme molecules
in contradiction to depletion theory.

  We find experimentally $\lim_{c_2 \rightarrow 0}\partial
\mbox{T}_{\mbox{\scriptsize{cloud}}} / \partial c_2$ = -0.15, 0.06
$\pm0.02$ $^{\circ}$C mg$^{-1}$ mL$^{-1}$ for PEG 1k and 8k
respectively. Using Eq.(\ref{eq:dTdp-virials}) with the measured
virial coefficients obtained independently from SLS yields
$\lim_{c_2 \rightarrow 0}\partial
\mbox{T}_{\mbox{\scriptsize{sp}}} /
\partial c_2$ = -0.03$\pm0.05$, 0.005$\pm0.008$ $^{\circ}$C mg$^{-1}$ mL$^{-1}$
for PEG 1k and 8k respectively. These results show that the
measured virial coefficients can correctly reproduce the sign of
$\lim_{\rho_2 \rightarrow 0}\partial
\mbox{T}_{\mbox{\scriptsize{cloud}}} /
\partial \rho_2$.
As discussed previously, $\gamma>0$ for a hard sphere mixture
whereas Fig.\ref{fig:lys-Tc-vs-PEG}
 shows that experimentally the sign of $\gamma$ varies
with PEG molecular weight. Therefore a depletion model cannot
account for the observed variation of the cloud point on polymer
concentration, but the measured virial coefficients in conjunction
with Eq.(\ref{eq:dTdp-virials}) do so qualitatively.

This paper demonstrates that the depletion theory does not
describe PEG/lysozyme mixtures. Firstly, the observation that
adding low molecular weight PEG depresses
T$_{\mbox{\scriptsize{cloud}}}$ whereas high molecular weight PEG
raises T$_{\mbox{\scriptsize{cloud}}}$ cannot be accounted for by
a pure depletion model. Secondly, depletion theory predicts that
adding PEG induces an attraction between lysozyme molecules
whereas light scattering revealed the opposite: PEG induces
repulsion between lysozyme molecules.  Lysozyme/PEG interactions
were characterized by virial coefficients obtained from light
scattering experiments. It is demonstrated that to even
qualitatively explain our data, the free energy must include third
virial terms.  The measured mixed virial coefficients are smaller
than those predicted for an equivalent hard sphere mixture and are
consistent with attractions between lysozyme and PEG. Models of
hydrogen bonding of water molecules to the PEG backbone
qualitatively explain PEG's phase behavior alone in water
\cite{bekiranov97}. Therefore we speculate that PEG can similarly
form hydrogen bonds with residues on the surface of lysozyme
molecules thereby creating the attraction between PEG and
lysozyme. The measured virial coefficients, combined with
thermodynamic theory predict the observed behavior of the cloud
point demonstrating the consistency of these two independent sets
of experiments. Therefore, an accurate model of the phase behavior
and light scattering of lysozyme/PEG mixtures must account for
both the entropic depletion effect and an energetic attraction
between protein and polymer.

\begin{acknowledgments}
Research was funded by the NASA Office of Biological \& Physical
Research, Fundamental Microgravity Research in Physical Sciences
(Fluid Physics) Grant \# NAG3-2386. G.M.T. was supported by the
National Institutes of Health grant \# NIH EY 11840.
\end{acknowledgments}

\begin{table*}
\begin{tabular}{|c|c|c|c|c|c|c|c|c|} \hline  & $M_2$ & $N$ &
$B_{22}$  & $B_{22}/B_{22}^{\mbox{\scriptsize{HS}}}$& $B_{12}$&
$B_{12}/B_{12}^{\mbox{\scriptsize{HS}}}$ &
$C_{112}$ & $C_{112}/C_{112}^{\mbox{\scriptsize{HS}}}$ \\
& [10$^3$g/mol] & & [10$^{-3}$mL mol/g$^2$] & & [10$^{-4}$mL
mol/g$^2$]& & [10$^{-4}$mL$^2$ mol/g$^3$] & \\
\hline PEG1k & 0.97 $\pm$ 0.1 & 22 &  9.2$\pm$ 1.0 & 0.5 &
12.9$\pm$1.2
& $ 0.34$ &  16.1$\pm$1.3 & $ 0.23 $ \\
\hline PEG8k & 10.4$\pm$0.5 & 236 &  3.35$\pm$0.12 & 0.4 &
1.7$\pm$0.3 &
$ 0.089 $ & 3.13$\pm$0.2 & $ 0.051 $ \\
\hline \end{tabular}
 \caption{PEG molecular weights ($M_2$) and
second virial coefficients ($B_{22}$) taken from light scattering
data at T$=30^{\circ}$C, the degree of PEG polymerization
($N=M_2/44$) and mixed virial coefficients ($B_{12}$,$C_{112}$)
for lysozyme/PEG solutions obtained from fits to $\alpha$ and
$\beta$, Eqs.(\ref{eq:2comp-Int},\ref{eq:2comp-Slope}), are shown.
The equivalent hard sphere values were obtained by using
$r_{\mbox{\scriptsize{g}}}$ of PEG to determine
$B_{22}^{\mbox{\scriptsize{HS}}}$ and
$r_{\mbox{\scriptsize{eff}}}$ of PEG along with
$r_{\mbox{\scriptsize{H}}}$ of lysozyme to determine
$B_{12}^{\mbox{\scriptsize{HS}}}$ and
$C_{112}^{\mbox{\scriptsize{HS}}}$ by
Eqs.(\ref{eq:secondvirials-hs},\ref{eq:thirdvirials-hs}). The
error bars came from the linear least squares fits.  }
\label{tab:cross-virials}
\end{table*}

\end{document}